# Time-series forecasting for nonlinear high-dimensional system using hybrid method combining autoencoder and multi-parallelized quantum long short-term memory and gated recurrent unit

Makoto Takagi[1], Ryuji Kokubo[1], Misato Kurosawa[1], Tsubasa Ikami[2], Yasuhiro Egami[3], Hiroki Nagai[2], Takahiro Kashikawa[4], Koichi Kimura[4], Yutaka Takita[4], Yu Matsuda[1, *]

1. Department of Modern Mechanical Engineering, Waseda University, 3-4-1 Ookubo, Shinjuku-ku, Tokyo, 169-8555, Japan

2. Institute of Fluid Science, Tohoku University, 2-1-1 Katahira, Aoba-ku, Sendai, Miyagi-prefecture 980-8577, Japan

3. Department of Mechanical Engineering, Aichi Institute of Technology, 1247 Yachigusa, Yakusa-Cho, Toyota, Aichi-prefecture 470-0392, Japan

4. Quantum Application Core Project, Quantum Laboratory, Fujitsu Research, Fujistu Ltd, Kawasaki, Kanagawa 211-8588, Japan

* Corresponding author: Yu Matsuda

Email: y.matsuda@waseda.jp

**Abstract**

A time-series forecasting method for high-dimensional spatial data is proposed. The method involves optimal selection of sparse sensor positions to efficiently represent the spatial domain, time-series forecasting at these positions, and estimation of the entire spatial distribution from the forecasted values via a learned decoder. Sensor positions are selected using a method based on combinatorial optimization. Introducing multi-parallelized quantum long short-term memory (MP-QLSTM) and gated recurrent unit (MP-QGRU) improves time-series forecasting performance by extending QLSTM models using the same number of variational quantum circuits (VQCs) as the cell state dimensions. Unlike the original QLSTM, our method fully measures all qubits in each VQC, maximizing the representation capacity. MP-QLSTM and MP-QGRU achieve approximately 1.5% lower test loss than classical LSTM and GRU. The root mean squared percentage error of MP-QLSTM is 0.256% against the values measured independently using semiconductor pressure sensors, demonstrating the method's accuracy and effectiveness for high-dimensional forecasting tasks.

**Main**

Time-series forecasting plays a crucial role in various fields of science, engineering, medicine, and social science. [1-7] A variety of forecasting methods have been proposed, ranging from conventional statistical methods to recent machine learning and deep learning approaches. Moving average, autoregressive (AR), autoregressive integrated moving average (ARIMA) models, Kalman filter and smoother, and particle (Monte Carlo) filter are well-known examples of statistical methods. [1,2,6,8] Support vector machine, Gaussian process, and classification and regression tree are examples of time-series forecasting by machine learning. [6,9-12] Advances in data storage and data acquisition technologies have led to the collection of large amounts of data, and the ability to process these data with significantly improving computing power has triggered a growing amount of research on time-series forecasting using deep learning approaches. Deep learning approaches excel in leveraging large amounts of time-series data to discover latent or/and complex relationships between multiple time series. In particular, approaches based on methods such as multilayer perceptrons (MLPs), recurrent neural networks (RNNs), and convolutional neural networks (CNNs) have been intensively investigated. [4-6,9,13-15] MLPs are simple feedforward networks and were one of the commonly used approaches. According to the universal approximation theorem, [16] MLPs are theoretically capable of approximating any continuous function given sufficient neurons and layers. In practice, however, they often struggle to efficiently approximate complex patterns in high-dimensional or structured data (for example, time-series data and images) due to challenges in training and optimization. [4,6,13,17] CNNs are equipped with convolutional layers that excel at extracting correlations and repeated patterns in data. [6,13,18,19] RNNs are designed to store information from the past, including trends and tendencies through their recurrent structures that connect the output of neurons to their inputs. [6,13,20,21] Long-short term memory (LSTM), which addresses the vanishing gradient problem of RNNs, is considered a promising method, and the addition of the forget gate further enhances its ability to effectively manage both long-term and short-term dependencies. [22-30] Gated recurrent unit (GRU) is used as a time-series forecasting method with a simpler architecture than LSTM to reduce the computational cost for learning. [31,32] LSTM and GRU play a central role in various time-series forecasting applications, such as healthcare monitoring systems, [33-35] weather forecasting, [36-39] energy consumption prediction, [40-45] inventory management and demand forecasting, [46-50] and financial data analytics. [4,51-55] Recently, Transformer models [56] have gained attention in time-series forecasting, [57] especially for data with long-term dependencies or extremely large datasets. However, LSTM and GRU are more advantageous in cases where short-term dependencies are crucial or the record length of measurement data is often still limited for high-resolution images and data acquired by sensor arrays. [58,59]

In the meantime, quantum computers have received much attention as has the development of noisy intermediate-scale quantum (NISQ) devices and the expansion of cloud services that enable access to them. [60-68] This has naturally led to a growing interest in quantum machine learning, which promises to revolutionize data analysis and computation. [69-73] By leveraging quantum parallelism and the expanded capabilities of quantum states, richer and more complex representations are expected, which has led proposals for the integration of variational quantum circuits (VQCs) [69,74,75] into LSTM architectures, referred to as quantum LSTM (QLSTM). [76-79] QLSTM has been reported to offer several advantages, such as faster convergence, higher representational power, and better prediction accuracy compared to classical LSTM models. [76-79] The architectures of QLSTM and VQCs within QLSTM have remained mostly unchanged since their original proposal, despite the proposal of a linear-layer-enhanced QLSTM (L-QLSTM) [78] that aims to improve learning performance, which leaves room for further research on the architectures of QLSTM and VQC to enhance learning capability. It is also expected that quantum GRU (QGRU) can be introduced by incorporating VQCs into GRU architecture to improve the capability of GRU.

While the focus has traditionally been on one- or low-dimensional time series data, high-dimensional spatial time-series data are increasingly being used, such as time-series images and data acquired by sensor arrays. These data are becoming increasingly important in fields such as analyzing unsteady flow fields around/inside objects, [80-83] seismic wave propagation, [84-86] monitoring land use and crop growth from remote sensing data, [87-90] and monitoring cell growth. [91-93] As processing large volumes of data is challenging, approaches such as reduced-order modeling and compressive sensing have been proposed. [94-101] These methods take advantage of the fact that high-dimensional systems can often be represented in low-dimensional forms by using tailored or data-driven bases, such as proper orthogonal decomposition (POD) [102,103] and dynamic mode decomposition. [104,105] Although time-series forecasting methods for such high-dimensional spatial time-series data are needed in these applications, such methods have not been proposed so far. This is because LSTMs, QLSTMs, and QGRU require high computational costs when applied to high-dimensional data.

In this study, a time-series forecasting method for high-dimensional spatial time-series data was proposed, as shown in Fig. 1. The proposed method consists of three steps: first, sparse sensor positions are extracted from time-series spatial data based on the optimal sensor placement problem. [96,106,107] In this process, the optimal sensor positions are determined to effectively represent the data and to perform sparse sensing based on reduced-order modeling, using an annealing (Ising) machine. [107] Second, time-series forecasting is performed at the selected sensor positions. In this study, multi-parallelized QLSTM (MP-QLSTM) and multi-parallelized QGRU (MP-QGRU) were proposed, which are extensions of QLSTM and QGRU, for time-series forecasting. Their performance was compared with that of LSTM, GRU, and QLSTM. Finally, based on the time-series forecasted data at the selected sensor positions, the entire spatial data is estimated using the learned decoder pre-trained on the data. The estimated data is compared with the data measured independently using semiconductor pressure sensors, demonstrating the effectiveness for high-dimensional forecasting tasks.

## Results
### Selection of optimal positions
Our forecasting method was applied to the flow measurement dataset of the Kármán vortex street. [107-109] In this study, five optimal sensor positions were selected by solving the optimal sensor placement problem with an annealing machine [107] as shown in Fig. 2. These selected positions are located in the region of the Kármán vortex street. In this study, the number of sensor positions was much smaller than in our previous study because a larger number of selected points leads to higher computational costs in the subsequent forecasting process.

### Time-series forecasting for selected optimal positions
The time-series forecasting was performed at these selected sensor positions using LSTM, GRU, L-QLSTM, MP-QLSTM, and MP-QGRU, respectively. The training and validation losses for LSTM, L-QLSTM, MP-QLSTM, and MP-QGRU are shown in Fig. 3, while the result of GRU can be found in the Supplementary information in Note 1. In the figure, the vertical axis represents the mean squared error (MSE). The training loss approaches a plateau around 20 epochs, and since the validation loss also stabilizes, the training can be considered complete at this point. While all methods show some fluctuations in validation loss, MP-QLSTM and MP-QGRU show more stable behavior, with gentler fluctuations and a gradual decrease in loss. It is noteworthy that the proposed method shows less fluctuation even though the same learning rate and batch size were used for all methods, the proposed method shows less fluctuation. For L-QLSTM, both test and validation losses showed a temporary increase around the 50th epoch. MP-QLSTM and MP-QGRU achieved a lower test loss compared to LSTM and L-QLSTM, and the difference between test and validation losses is smaller for LSTM and L-QLSTM. These observations indicate that MP-QLSTM and MP-QGRU provide more stable and consistent learning. The test losses (MSEs) for LSTM, GRU, L-QLSTM, MP-QLSTM, and MP-QGRU were $6.49 \times 10^{-2}$, $6.47 \times 10^{-2}$, $7.36 \times 10^{-2}$, $6.40 \times 10^{-2}$, and $6.39 \times 10^{-2}$, respectively. The test losses of MP-QLSTM and MP-QGRU were approximately 1.5% lower than those of LSTM and GRU and approximately 15% lower than that of L-QLSTM. Although the VQC in MP-QLSTM used only three qubits, less than the four qubits in L-QLSTM, MP-QLSTM provided better results. The time-series forecasting results of LSTM, MP-QLSTM, and MP-QGRU are shown in Fig. 4. As shown in the figure, there are no significant differences between the methods. It is worth noting that LSTM tends to underestimate local minima, especially around time step 60. The results for the other methods can be found in the Supplementary information in Note 2.

**Estimation of high-dimensional spatial distribution**
The entire pressure distributions were predicted from these time-forecasting data at the five selected sensor positions. As representative examples predicted by LSTM and MP-QLSTM, Fig. 5 presents the entire pressure distributions predicted using the decoder part of an autoencoder. Although the fine structures of the flow field were not accurately predicted, both methods have successfully captured the overall structure, despite relying on data from only five sensor positions. To quantitatively evaluate the validity of the predicted pressure distributions, the root mean squared percentage error (RMSPE) was calculated using the pressure data obtained from the semiconductor pressure sensor via the pressure tap. The RMSPE for LSTM was 0.264%, while the RMSPE for MP-QLSTM was 0.256%, indicating that MP-QLSTM performed better results.

To summarize, the series of methods proposed in this study—selecting a small number of sensor positions that efficiently represent the entire distribution from the dataset, performing time-series forecasting at these positions using MP-QLSTM or MP-QGRU, and estimating the entire distribution at the next time step using a decoder based on the predicted data—has been shown to be effective.

To validate the effectiveness on different data, we also investigated the performance of MP-QLSTM and MP using solar power data provided by the National Renewable Energy Laboratory (NREL). [110] As a result, the test loss of MP-QLSTM was approximately 3% lower than those of LSTM and GRU and approximately 10% lower than that of L-QLSTM, demonstrating its effectiveness in forecasting nonlinear data. The details of the results can be found in the Supplementary information in Note 3. A more detailed comparison between MP-QLSTM and LSTM was conducted using data generated by the Lorenz equations with added Gaussian noise. As a result, MP-QLSTM is more suitable for forecasting data with high-noise. On the other hand, when the noise level is low, LSTM performs better, suggesting that it is important to choose between the two methods depending on the characteristics of the data. The details of the results can be found in the Supplementary information in Note 4.

**Discussion**
We propose a time-series forecasting method for high-dimensional spatial time-series data. The method consists of three steps: selection of sparse sensor positions for efficient representation of high-dimensional spatial data, time-series forecasting at selected sensor positions, and estimation of high-dimensional spatial data from the forecasted data. The optimal sensor positions are determined using our previously proposed method using an annealing (Ising) machine.

As a time-series forecasting method, we propose multi-parallelized quantum long short-term memory (MP-QLSTM) and gated recurrent unit (MP-QGRU). QLSTM was originally introduced by replacing classical neural networks in LSTM with variational quantum circuits (VQCs). However, to match the dimensions of the hidden layer and the cell state, only a subset of the qubits is measured, which results in only partially using of the expressive power of the VQC. To address such issues in the original QLSTM, the same number of VQCs as the dimension of the cell states are used in MP-QLSTM and MP-QGRU, where all qubits are measured. The test losses of MP-QLSTM and MP-QGRU were approximately 1.5% lower than those of LSTM and GRU and approximately 15% lower than that of L-QLSTM. Considering that LSTM is a highly optimized baseline, even a 1% improvement in test loss is not trivial. Although the VQC in MP-QLSTM used only three qubits, less than the four qubits used in L-QLSTM, MP-QLSTM yielded better results.

Based on the forecasted data at five selected sensor points, the entire high-dimensional (780 × 780) spatial distribution is estimated by a decoder which is learned by an autoencoder. To quantitatively evaluate the validity of the predicted distributions, the root mean squared percentage error (RMSPE) is calculated using the pressure data separately obtained by the semiconductor pressure sensor. The RMSPE for MP-QLSTM was 0.256%, indicating that our method demonstrates good agreement even though it was estimated from the data at only five points.

## Methods
### Proposed method for time-series forecasting of high-dimensional data
The proposed method consists of three steps: selection of sparse sensor positions to efficiently represent high-dimensional spatial data, time-series forecasting at selected sensor positions, and estimation of high-dimensional spatial data from the forecasted data. In this study, sparse sensor positions were selected as a solution to the optimal sensor placement problem. [96,106,107] The optimal sensor placement problem was solved using our previously proposed method with an annealing machine, [107] and five sensor positions were selected. In the following sections, we introduce the time-series forecasting methods and, sequentially, the method estimating the entire spatial data from the forecasted data.

### Time-series forecasting using MP-QLSTM
As the second step of the proposed method, time-series forecasting is performed at the selected optimal positions. As mentioned in the introduction, QLSTM is one of the most promising methods for time-series forecasting using NISQ devices. A QLSTM is introduced by replacing classical neural networks in LSTM with VQCs, [76] which are quantum circuits with learnable parameters that are iteratively optimized. [69,74,75] In this original QLSTM, a number of qubits equal to the sum of the dimensions of input data and the hidden layer must be prepared. However, to match the dimensions of the hidden layer and the cell state, only a subset of the qubits is measured, resulting in the partial use of the expressive power of the VQC. To address such issues in the original QLSTM, a method called liner-layer-enhanced QLSTM (L-QLSTM) has been proposed, [78] which embeds linear layers before and after each VQC. In this approach, the linear layer before each VQC reduces the dimensions of the input data and hidden state, allowing flexibility in the number of qubits used in each VQC. At the same time, the dimension of the output from each VQC is adjusted to match that of the cell state using the linear layer after it. With L-QLSTM, all qubits in the VQCs can be measured, improving the ability to learn parameters. Although L-QLSTM overcomes the issue in the original QLSTM where not all qubits are measured, it introduces a new problem where the linear layer placed before the VQC reduces the information from the input data and hidden state. Then, we propose a novel method called multi-parallelized QLSTM (MP-QLSTM), which uses the same number of VQCs as the dimension of the cell states. The architecture of MP-QLSTM is depicted in Figure 6. In MP-QLSTM, $M$ linear layers are used to reduce the combined size of input and hidden states while preserving as much information as possible, adjusting it to match the number of qubits used in each VQC. The formulas for the output state from VQC used in MP-QLSTM are as follows:

$$\mathbf{v}_t = [\mathbf{h}_{t-1}, \mathbf{x}_t], \tag{1}$$

$$g_k^\alpha = \mathcal{Q}_k^\alpha \left( \mathcal{L}_{k,1}^\alpha(\mathbf{v}_t), \mathcal{L}_{k,2}^\alpha(\mathbf{v}_t), \cdots, \mathcal{L}_{k,M}^\alpha(\mathbf{v}_t) \right), \tag{2}$$

where subscript $t$ and $t-1$ represent the time step, $\mathbf{x}$ and $\mathbf{h}$ are the input and hidden states, respectively. The operator $[\mathbf{a}, \mathbf{b}]$ concatenates the vectors $\mathbf{a}$ and $\mathbf{b}$. The $\mathcal{Q}_k^\alpha$ represents the calculation at the $k$-th ($k = 1, 2, \cdots, K$) VQC for $\alpha$ gate, where $K$ is the number of VQCs used and equal to the dimension of the cell state, and $\alpha \in \{f, i, c, o\}$ where f, i, c, and o denote the forget gate, input gate, output gate, and cell state, respectively. The $\mathcal{L}_{k,m}^\alpha$ represents the $m$-th ($m = 1, 2, \cdots, M$) linear layer calculation of $\mathbf{v}_t$ for the $k$-th VQC of the $\alpha$ gate, where the number of linear layers, $M$, can be freely determined. Although it is possible to use MLPs instead of linear layers, the use of MLPs often leads to issues such as gradient explosion or vanishing gradients. Furthermore, as confirmed in the 'Results and Discussion' section, the MP-QLSTM using linear layers demonstrated sufficient performance. Therefore, we used linear layers in this study. The formulas for MP-QLSTM are as follows:

$$\mathbf{f}_t = \sigma([g_1^f, g_2^f, \cdots, g_K^f]), \tag{3}$$

$$\mathbf{i}_t = \sigma([g_1^i, g_2^i, \cdots, g_K^i]), \tag{4}$$

$$\tilde{\mathbf{c}}_t = \tanh([g_1^c, g_2^c, \cdots, g_K^c]), \tag{5}$$

$$\mathbf{o}_t = \sigma([g_1^o, g_2^o, \cdots, g_K^o]), \tag{6}$$

$$\mathbf{c}_t = \mathbf{f}_t \odot \mathbf{c}_{t-1} + \mathbf{i}_t \odot \tilde{\mathbf{c}}_t, \tag{7}$$

$$\mathbf{h}_t = \mathbf{o}_t \odot \tanh(\mathbf{c}_t), \tag{8}$$

where the operator $\odot$ represents the Hadamard product. The sigmoid and hyperbolic tangent functions are represented by $\sigma$ and tanh, respectively. The $\mathbf{f}$, $\mathbf{i}$, and $\mathbf{o}$ are respectively the forget, input, and output gates. The $\mathbf{c}$ represents cell states.

In MP-QLSTM, the same number of VOQs as the dimension of the cell state is used for each gate. A VQC consists of three components: encoding, variational, and measurement layers. In the encoding layer, classical data are encoded into corresponding quantum states from computational basis states $|0\rangle$ and $|1\rangle$ (usually $|0\rangle$ is used) by applying X- and Y-rotation gates, denoted as $R_X$ and $R_Y$, respectively. The notation $|\rangle$ is known as Dirac notation. To simplify the explanation, we describe the case with three linear layers $\mathcal{L}_{k,m}^\alpha$ ($m = 1, 2, 3$) and three qubits as illustrated in Fig. 7. In the encoding layer, three linear layers are used to reduce the combined size of input and hidden states to match the number of qubits (three), which enables better preservation of input and hidden state information compared to using a single linear layer. That is, the three outputs $z_{k,1,1}$, $z_{k,1,2}$, and $z_{k,1,3}$ from the first linear layer $\mathcal{L}_{k,1}^\alpha$ are encoded as the rotation angles of the $R_X$ gates for each qubit. Subsequently, the outputs from the second and third linear layers, $\mathcal{L}_{k,2}^\alpha$ and $\mathcal{L}_{k,3}^\alpha$, are sequentially encoded onto each qubit as the rotation angles of the $R_Y$ and $R_Z$ gates, respectively. In the variational layer, a quantum circuit with qubit operations, such as rotation and entanglement, is considered. The parameters of the operations are learnable. The entanglement of qubits is created using a combination of a controlled-NOT (CNOT) gate and a superposition state of computational basis states. In the variational layer used in Refs. [76,79], all rotation gates are placed after the CNOT gates. Then, we prepare a VQC where rotation gates are placed between entangling CNOT gates. In the measurement layer, multiple-qubit measurements by Pauli-$Z$ operators are conducted and the elements denoted as $g_k^\alpha$ are obtained for each VQC. By setting $M = 1$ in Eq. (2) (i.e., using a single linear layer) and replacing the VQC with the identity operator, MP-QLSTM becomes equivalent to a classical LSTM. Therefore, it can be considered as a generalized form of classical LSTM, which has the potential for performance improvement.

In this study, we also introduce an MP-QGRU. MP-QGRU is directly introduced from the classical GRU as the same manner as MP-QLSTM. The architecture of MP-QGRU is depicted in Fig. 8. The formulas for MP-QGRU are as follows:

$$\mathbf{r}_t = \sigma([g_1^r, g_2^r, \cdots, g_K^r]), \tag{9}$$

$$\mathbf{z}_t = \sigma([g_1^z, g_2^z, \cdots, g_K^z]), \tag{10}$$

$$\tilde{\mathbf{h}}_t = \tanh([g_1^h, g_2^h, \cdots, g_K^h]), \tag{11}$$

$$\mathbf{h}_t = (1 - \mathbf{z}_t) \odot \mathbf{h}_{t-1} + \mathbf{z}_t \odot \tilde{\mathbf{h}}_t, \tag{12}$$

where the $\mathbf{r}$ and $\mathbf{z}$ are the reset and update gates, respectively. The $g_k^r$ and $g_k^z$ represent the outputs from the $k$-th ($k = 1, 2, \cdots, K$) VQC for reset and update gates as the same manner as Eqs. (1) and (2) ($\mathbf{v}_t$ and $g_k^\alpha$), respectively. Here, $K$ is the number of VQCs used and equal to the dimension of the hidden state. The output $g_k^h$ from the $k$-th VQC for $\tilde{\mathbf{h}}_t$ is calculated as follows:

$$\tilde{\mathbf{v}}_t = [\mathbf{x}_t, \mathbf{r}_t \odot \mathbf{h}_{t-1}], \tag{13}$$

$$g_k^h = \mathcal{Q}_k^h\left(\mathcal{L}_{k,1}^h(\tilde{\mathbf{v}}_t), \mathcal{L}_{k,2}^h(\tilde{\mathbf{v}}_t), \cdots, \mathcal{L}_{k,M}^h(\tilde{\mathbf{v}}_t)\right). \tag{14}$$

In this study, VQCs with the same architecture as those used for MP-QLSTM were used.

**Estimation of high-dimensional data from forecasted data at selected position**
Based on the time-series forecasted data at selected positions, the entire spatial data is estimated using MLP. The following autoencoder is considered:

$$\mathbf{z} = \mathbf{f}_{\text{enc}}(\mathbf{d}; \mathbf{\psi}_{\text{enc}}), \tag{15}$$

$$\hat{\mathbf{d}} = \mathbf{f}_{\text{dec}}(\mathbf{z}; \mathbf{\psi}_{\text{dec}}), \tag{16}$$

where $\mathbf{d}$ and $\hat{\mathbf{d}}$ are input and output data (high-dimensional spatial time-series data), respectively. The state $\mathbf{z}$ is a latent state. The functions $\mathbf{f}_{\text{enc}}$ and $\mathbf{f}_{\text{dec}}$ represent encoder and decoder, respectively. The learnable parameters are represented as $\mathbf{\psi}_{\text{enc}}$ and $\mathbf{\psi}_{\text{dec}}$. Here, we consider the following MLP, $\mathbf{f}_{\text{trans}}$, that transforms the data at the selected optimal positions, $\mathbf{d}_s$, into another latent variable, $\mathbf{z}_{\text{trans}}$, corresponding to $\mathbf{z}$.

$$\mathbf{z}_{\text{trans}} = \mathbf{f}_{\text{trans}}(\mathbf{d}_s; \mathbf{\psi}_{\text{trans}}), \tag{17}$$

where $\mathbf{\psi}_{\text{trans}}$ is a learnable parameter vector. Then, the entire spatial data, $\mathbf{d}_{\text{ent}}$, is estimated as

$$\mathbf{d}_{\text{ent}} = \mathbf{f}_{\text{dec}}(\mathbf{z}_{\text{trans}}; \boldsymbol{\psi}_{\text{dec}}), \tag{18}$$

where $\mathbf{f}_{\text{dec}}$ and $\boldsymbol{\psi}_{\text{dec}}$ are the same as those in Eq. (16). The parameters $\boldsymbol{\psi}_{\text{enc}}$, $\boldsymbol{\psi}_{\text{dec}}$, and $\boldsymbol{\psi}_{\text{trans}}$ are learned by multi-task learning [111,112] considering the following loss function, $L_{\text{edt}}$:

$$L_{\text{edt}}(\boldsymbol{\psi}_{\text{enc}}, \boldsymbol{\psi}_{\text{dec}}, \boldsymbol{\psi}_{\text{trans}}) = \left\| \mathbf{d} - \hat{\mathbf{d}} \right\|_2^2 + \left\| \mathbf{d} - \mathbf{d}_{\text{ent}} \right\|_2^2, \tag{19}$$

where $\|\mathbf{a}\|_2$ represents the $\ell 2$ norm of the vector $\mathbf{a}$. With the learned parameters, we can estimate an entire spatial data from the forecasted data at the optimal sensor positions.

**High-dimensional spatial time-series data for demonstration**

The flow image data obtained in our previous studies [107-109] was used as high-dimensional spatial time-series data, which were measured using the pressure-sensitive paint (PSP) method. [83,113,114] The time-series spatial data used were the pressure distribution induced by the Kármán vortex street behind a square cylinder (width: 40 mm, length: 40 mm, and height: width: 100 mm) with main flow velocity of 50 m/s and the Reynolds number of $1.1 \times 10^5$. The data had a spatial grid of $780 \times 780$ and 8192 time point. This pressure distribution data, initially formatted as a two-dimensional array, was reshaped into a one-dimensional array and arranged temporally to create a data matrix as $\mathbf{P} = [\mathbf{p}_1 \ \mathbf{p}_2 \ \cdots \ \mathbf{p}_N]$, where $\mathbf{p}_n$ represents the $n$-th reshaped pressure data. We considered the singular value decomposition (SVD) of $\mathbf{P}$ as

$$\mathbf{P} = \mathbf{U}\boldsymbol{\Sigma}\mathbf{V}^\top, \tag{20}$$

where the matrices $\mathbf{U}$ and $\mathbf{V}$ are unitary matrices, respectively. The superscript ⊤ represents the transpose. The matrix $\boldsymbol{\Sigma}$ is a diagonal matrix in which the singular values are sorted in descending order. Each column of the matrix $\mathbf{U}$ represents the proper orthogonal decomposition (POD) mode as $\mathbf{U} = [\mathbf{u}_1 \ \mathbf{u}_2 \ \cdots \ \mathbf{u}_n]$. The optimal sensor placement problem is considered based on the POD modes. [96,106,107] It is well known that noise is contained in the higher-order singular values in SVD, [115,116] and the noise reduction data $\widetilde{\mathbf{P}}$ can be obtained by considering a truncated SVD as follows:

$$\widetilde{\mathbf{P}} = \widetilde{\mathbf{U}}\widetilde{\boldsymbol{\Sigma}}\widetilde{\mathbf{V}}^\top, \tag{21}$$

where $\widetilde{\boldsymbol{\Sigma}}$ is an $r \times r$ diagonal matrix containing the leading $r$ singular values. The matrices $\widetilde{\mathbf{U}}$ and $\widetilde{\mathbf{V}}$ are the corresponding reduced matrices. Although several sophisticated noise reduction methods have been proposed, [109,117,118] we employed a simple truncated SVD method [116] with truncated rank $r = 20$ to focus on time-series forecasting, and the denoised time-series spatial data is used to evaluate the performance of MP-QLSTM. Moreover, for the comparison, the pressures were independently measured through pressure taps using pressure transducers, separately from the PSP method.

In the following, we provide a detailed explanation of the data processing. From the 8192 time points, 70% were used as training data, and the following 20% as validation data, and the remaining 10 % as the test data. Five sensor positions were selected based on the calculated POD modes using Fujitsu computing as a service digital annealer and were illustrated in Figure 2. The details of the calculations are provided in our previous paper. [107]

Subsequently, the time-series forecasting was performed for the data at the five selected positions. In this process, the performance of MP-QLSTM and MP-QGRU was compared with that of LSTM and L-QLSTM. As the performance of the original QLSTM has been reported to be worse than that of LSTM and L-QLSTM, [78] this study focused on the comparison between LSTM, L-QLSTM, and our proposed MP-QLSTM and MP-QGRU. In this study, we implemented the calculations through a script developed using Qiskit [119,120] and PyTorch. [121,122] In the calculation of MP-QLSTM and MP-QGRU, the data at the five selected positions were used as the input state $\mathbf{x}$, with data from four consecutive time steps serving as the sequential input to predict the output at the next time step. The loss function was defined as the $\ell 2$ norm of the difference between the predicted output and the ground truth at the corresponding time step. The dimension of the hidden state $\mathbf{h}$, the number of linear layers for the VQC, $M$, and the number of qubits for each VQC were set to 5, 3, and 3, respectively. The training was performed using the Adaptive moment estimation (Adam) optimizer of "torch.optim" with a learning rate of 0.02, and the batch size was set to 128. In the calculation of LSTM, GRU, and L-QLSTM, the data at the selected five positions were also used as the input state, and the dimension of hidden states was also five. Four consecutive data points were used as sequential input to predict the output at the next time step. In the L-QLSTM calculation, the number of qubits was set to 4 in VQCs, following the reference, [78] which is larger than the number used in MP-QLSTM and MP-QGRU.

In the parameter learning of the autoencoder, we used three-layer MLPs for $\mathbf{f}_{\text{enc}}$ and $\mathbf{f}_{\text{dec}}$, and eight-layer MLPs for $\mathbf{f}_{\text{trans}}$, respectively, with the rectified linear unit (ReLU) as the activation function. In the encoder $\mathbf{f}_{\text{enc}}$, the dimension of the reshaped pressure data was reduced to 1,000 dimensions in the first layer, to 500 dimensions in the next layer, and then successively to 128 dimensions. In the decoder $\mathbf{f}_{\text{dec}}$, the process is reversed, where the 128-dimensional data is restored to the original dimensions. In the transformer $\mathbf{f}_{\text{trans}}$, the first layer increased the number of dimensions from 5, corresponding to the number of sensor positions, to 128, and the number of dimensions of subsequent layers was maintained at 128. The training was conducted using the Adam optimizer with a batch size of 32.

## Data availability
The flow measurement dataset obtained by the PSP method is available in Zenodo with identifier: https://doi.org/10.5281/zenodo.10215642.

## Code availability
The code for the time-series clustering developed in this study is included in "Supplemental information."

## Acknowledgments

Part of the work was conducted under the Collaborative Research Project J23I041, J24I064, and J25I033 of the Institute of Fluid Science, Tohoku University.  Finally, we would like to thank Editage (www.editage.com) for English language editing.

**Figures and Tables**

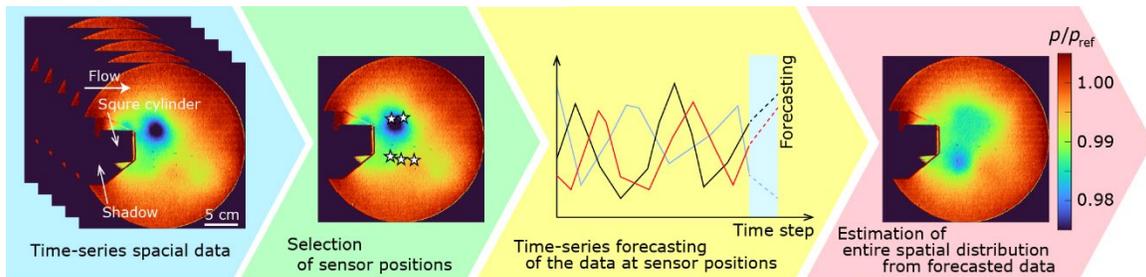

**Figure 1** Procedure of time-series forecasting method for high-dimensional spatial time-series data. First, the optimal sparse sensor positions are selected to efficiently represent the spatial domain. Second, time-series forecasting at these positions are performed. Finally, the entire spatial distribution is estimated from the forecasted values via a learned decoder.

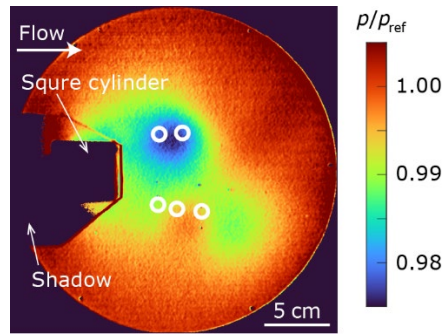

**Figure 2** Selected sensor positions. The sensor positions are estimated by our previous study using an annealing machine and are overlaid on a pressure distribution taken from a certain time step.

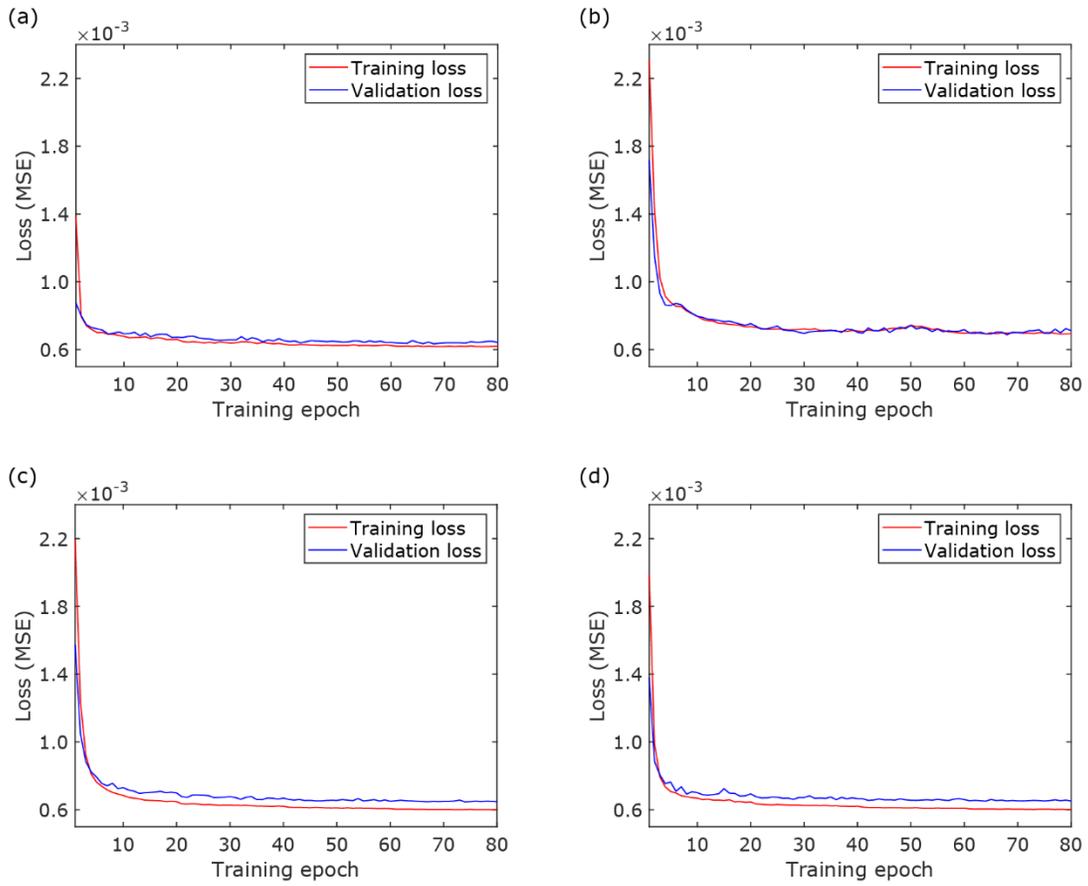

**Figure 3** Training and validation loss for pressure data. Results of (a) LSTM, (b) L-QLSTM, (c) MP-QLSTM, and (d) MP-QGRU.

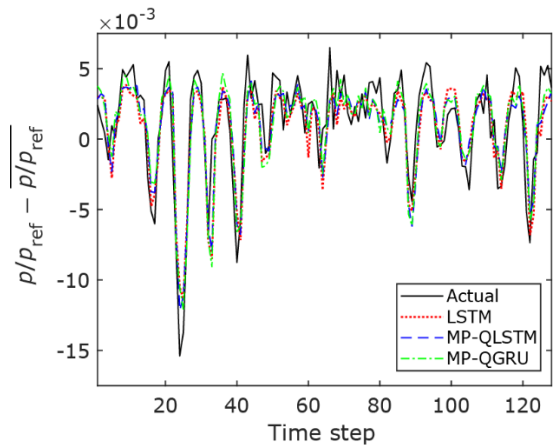

**Figure 4** Results of time-series forecasting for pressure data. The forecasting results of LSTM, MP-QLSTM, and MP-QGRU are shown.
11

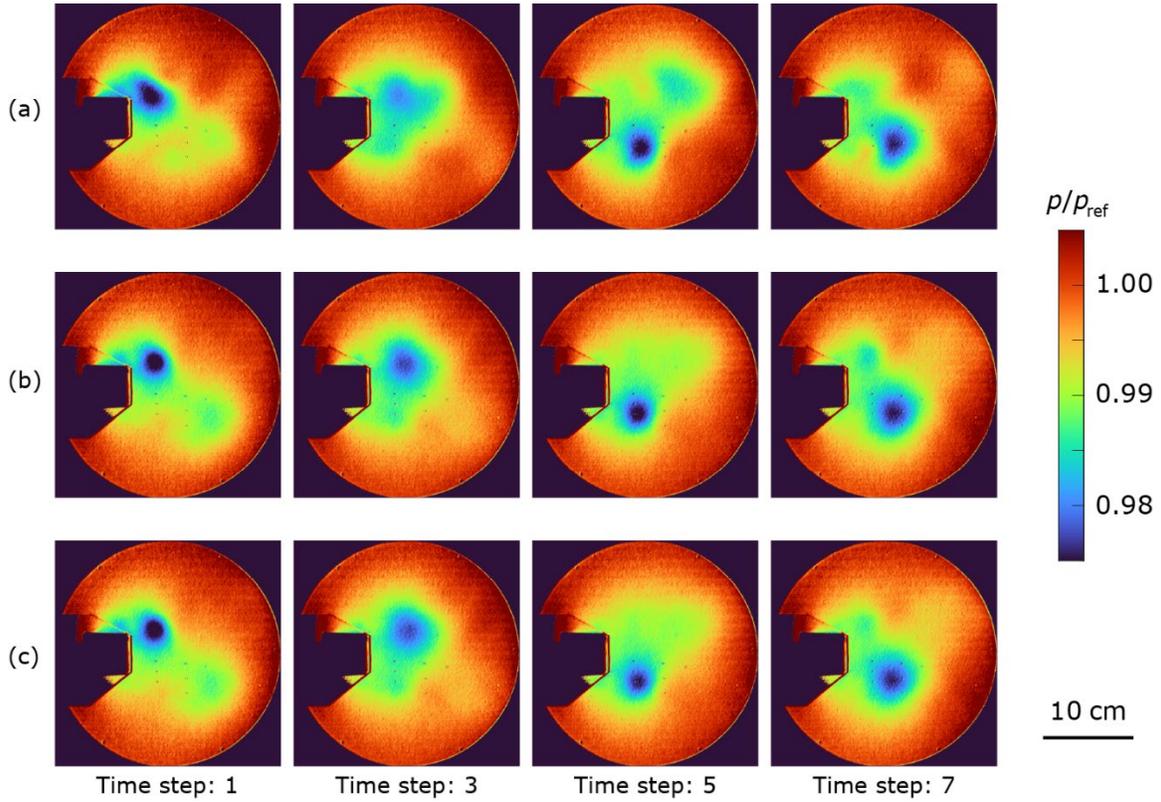

**Figure 5** Estimated entire pressure distributions. The distributions are estimated by using the decoder part of an autoencoder based on the data predicted by (a) Actual, (b) LSTM, and (c) MP-QLSTM. The pressure $p$ is normalized by an atmospheric pressure $p_{\text{ref}}$.

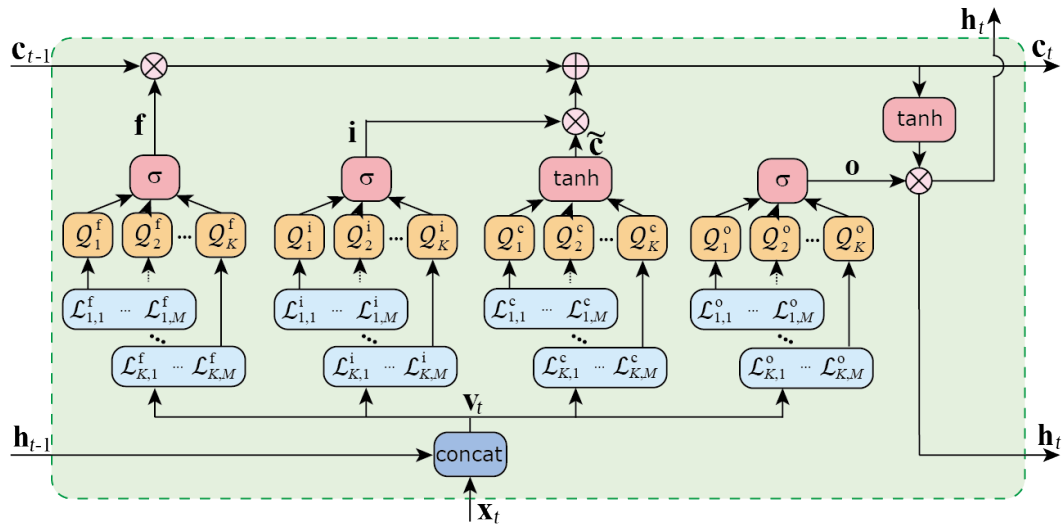

**Figure 6** Architecture of MP-QLSTM. The number of VQCs for each gate is equal to the dimension of cell state.

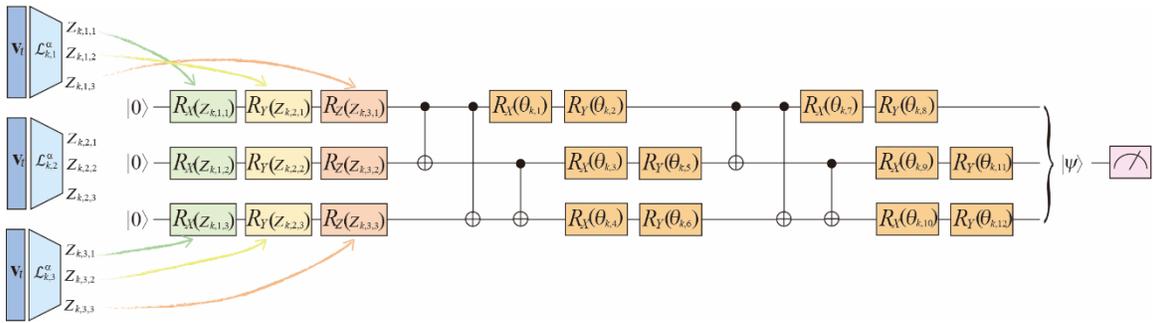

**Figure 7** Architecture of VQC. A VQC having three qubits is shown as a typical example.

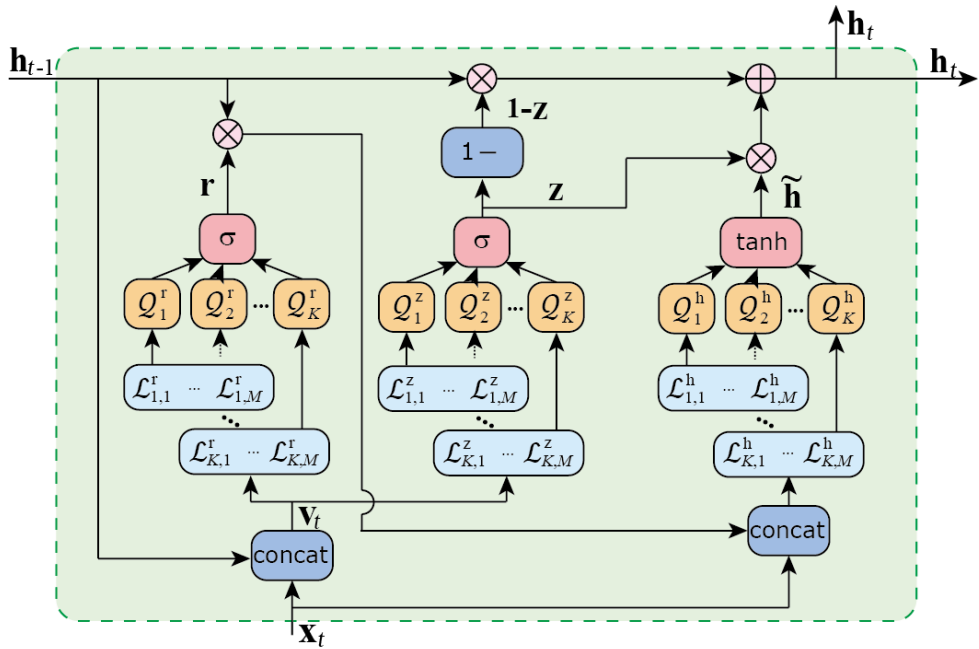

**Figure 8** Architecture of MP-QGRU. The number of VQCs for each gate is equal to the dimension of the cell state.

**References**

1 Chatfield, C. *Time-Series Forecasting*. (CRC Press, 2000).
2 Box, G. E. P., Jenkins, G. M., Reinsel, G. C. & Ljung, G. M. *Time Series Analysis: Forecasting and Control*. (Wiley, 2015).
3 Deb, C., Zhang, F., Yang, J., Lee, S. E. & Shah, K. W. A review on time series forecasting techniques for building energy consumption. *Renewable and Sustainable Energy Reviews* **74**, 902-924 (2017). https://doi.org/https://doi.org/10.1016/j.rser.2017.02.085
4 Sezer, O. B., Gudelek, M. U. & Ozbayoglu, A. M. Financial time series forecasting with deep learning : A systematic literature review: 2005–2019. *Applied Soft Computing* **90**, 106181 (2020). https://doi.org/https://doi.org/10.1016/j.asoc.2020.106181
5 Lim, B. & Zohren, S. Time-series forecasting with deep learning: a survey. *Philosophical Transactions of the Royal Society A: Mathematical, Physical and Engineering Sciences* **379**, 20200209 (2021). https://doi.org/doi:10.1098/rsta.2020.0209
6 Petropoulos, F. *et al.* Forecasting: theory and practice. *International Journal of Forecasting* **38**, 705-871 (2022). https://doi.org/https://doi.org/10.1016/j.ijforecast.2021.11.001
7 Martin, G. M. *et al.* Bayesian forecasting in economics and finance: A modern review. *International Journal of Forecasting* **40**, 811-839 (2024). https://doi.org/https://doi.org/10.1016/j.ijforecast.2023.05.002
8 Kitagawa, G. *Introduction to Time Series Modeling*. (CRC Press, Boca Raton 2010).
9 Makridakis, S., Spiliotis, E. & Assimakopoulos, V. Statistical and Machine Learning forecasting methods: Concerns and ways forward. *PLOS ONE* **13**, e0194889 (2018). https://doi.org/10.1371/journal.pone.0194889
10 Sapankevych, N. I. & Sankar, R. Time Series Prediction Using Support Vector Machines: A Survey. *IEEE Computational Intelligence Magazine* **4**, 24-38 (2009). https://doi.org/10.1109/MCI.2009.932254
11 Cheng, C., Sa-Ngasoongsong, A., Beyca, O., Le, T., Yang, H., Kong, Z. & Bukkapatnam, S. T. S. Time series forecasting for nonlinear and non-stationary processes: a review and comparative study. *IIE Transactions* **47**, 1053-1071 (2015). https://doi.org/10.1080/0740817X.2014.999180
12 Liu, H., Ong, Y. S., Shen, X. & Cai, J. When Gaussian Process Meets Big Data: A Review of Scalable GPs. *IEEE Transactions on Neural Networks and Learning Systems* **31**, 4405-4423 (2020). https://doi.org/10.1109/TNNLS.2019.2957109
13 Han, Z., Zhao, J., Leung, H., Ma, K. F. & Wang, W. A Review of Deep Learning Models for Time Series Prediction. *IEEE Sens. J.* **21**, 7833-7848 (2021). https://doi.org/10.1109/JSEN.2019.2923982
14 Torres, J. F., Hadjout, D., Sebaa, A., Martínez-Álvarez, F. & Troncoso, A. Deep Learning for Time Series Forecasting: A Survey. *Big Data* **9**, 3-21 (2021). https://doi.org/10.1089/big.2020.0159
15 Lara-Benítez, P., Carranza-García, M. & Riquelme, J. C. An Experimental Review on Deep Learning Architectures for Time Series Forecasting. *International Journal of Neural Systems* **31**, 2130001 (2021). https://doi.org/10.1142/s0129065721300011
16 Hornik, K., Stinchcombe, M. & White, H. Multilayer feedforward networks are universal approximators. *Neural Networks* **2**, 359-366 (1989). https://doi.org/https://doi.org/10.1016/0893-6080(89)90020-8
17 Bataineh, A. A., Kaur, D. & Jalali, S. M. J. Multi-Layer Perceptron Training Optimization Using Nature Inspired Computing. *IEEE Access* **10**, 36963-36977 (2022). https://doi.org/10.1109/ACCESS.2022.3164669
18 Borovykh, A., Bohte, S. & Oosterlee, C. W. Conditional Time Series Forecasting with Convolutional Neural Networks. arXiv:1703.04691 (2017). <https://ui.adsabs.harvard.edu/abs/2017arXiv170304691B>.
19 Koprinska, I., Wu, D. & Wang, Z. Convolutional Neural Networks for Energy Time Series Forecasting. in *2018 International Joint Conference on Neural Networks (IJCNN)*. 1-8. https://doi.org/10.1109/IJCNN.2018.8489399

20    Hewamalage, H., Bergmeir, C. & Bandara, K. Recurrent Neural Networks for Time Series Forecasting: Current status and future directions. *International Journal of Forecasting* **37**, 388-427 (2021). https://doi.org/https://doi.org/10.1016/j.ijforecast.2020.06.008
21    Connor, J. T., Martin, R. D. & Atlas, L. E. Recurrent neural networks and robust time series prediction. *IEEE Transactions on Neural Networks* **5**, 240-254 (1994). https://doi.org/10.1109/72.279188
22    Hochreiter, S. & Schmidhuber, J. Long Short-Term Memory. *Neural Computation* **9**, 1735-1780 (1997). https://doi.org/10.1162/neco.1997.9.8.1735
23    Gers, F. A., Schmidhuber, J. & Cummins, F. Learning to forget: continual prediction with LSTM. in *1999 Ninth International Conference on Artificial Neural Networks ICANN 99. (Conf. Publ. No. 470).* 850-855 vol.852. https://doi.org/10.1049/cp:19991218
24    Gers, F. A. & Schmidhuber, J. Recurrent nets that time and count. in *Proceedings of the IEEE-INNS-ENNS International Joint Conference on Neural Networks. IJCNN 2000. Neural Computing: New Challenges and Perspectives for the New Millennium.* 189-194 vol.183. https://doi.org/10.1109/IJCNN.2000.861302
25    Bahdanau, D. Neural machine translation by jointly learning to align and translate. *arXiv preprint arXiv:1409.0473* (2014).
26    Greff, K., Srivastava, R. K., Koutník, J., Steunebrink, B. R. & Schmidhuber, J. LSTM: A Search Space Odyssey. *IEEE Transactions on Neural Networks and Learning Systems* **28**, 2222-2232 (2017). https://doi.org/10.1109/TNNLS.2016.2582924
27    Yu, Y., Si, X., Hu, C. & Zhang, J. A Review of Recurrent Neural Networks: LSTM Cells and Network Architectures. *Neural Computation* **31**, 1235-1270 (2019). https://doi.org/10.1162/neco_a_01199
28    Staudemeyer, R. C. & Morris, E. R. Understanding LSTM--a tutorial into long short-term memory recurrent neural networks. *arXiv preprint arXiv:1909.09586* (2019).
29    Smagulova, K. & James, A. P. A survey on LSTM memristive neural network architectures and applications. *The European Physical Journal Special Topics* **228**, 2313-2324 (2019). https://doi.org/10.1140/epjst/e2019-900046-x
30    Lindemann, B., Müller, T., Vietz, H., Jazdi, N. & Weyrich, M. A survey on long short-term memory networks for time series prediction. *Procedia CIRP* **99**, 650-655 (2021). https://doi.org/https://doi.org/10.1016/j.procir.2021.03.088
31    Cho, K., Van Merriënboer, B., Gulcehre, C., Bahdanau, D., Bougares, F., Schwenk, H. & Bengio, Y. Learning phrase representations using RNN encoder-decoder for statistical machine translation. *arXiv preprint arXiv:1406.1078* (2014).
32    Chung, J., Gulcehre, C., Cho, K. & Bengio, Y. Empirical evaluation of gated recurrent neural networks on sequence modeling. *arXiv preprint arXiv:1412.3555* (2014).
33    Miotto, R., Wang, F., Wang, S., Jiang, X. & Dudley, J. T. Deep learning for healthcare: review, opportunities and challenges. *Briefings in Bioinformatics* **19**, 1236-1246 (2017). https://doi.org/10.1093/bib/bbx044
34    Bolhasani, H., Mohseni, M. & Rahmani, A. M. Deep learning applications for IoT in health care: A systematic review. *Informatics in Medicine Unlocked* **23**, 100550 (2021). https://doi.org/https://doi.org/10.1016/j.imu.2021.100550
35    Morid, M. A., Sheng, O. R. L. & Dunbar, J. Time Series Prediction Using Deep Learning Methods in Healthcare. *ACM Trans. Manage. Inf. Syst.* **14**, Article 2 (2023). https://doi.org/10.1145/3531326
36    Qing, X. & Niu, Y. Hourly day-ahead solar irradiance prediction using weather forecasts by LSTM. *Energy* **148**, 461-468 (2018). https://doi.org/https://doi.org/10.1016/j.energy.2018.01.177
37    Hossain, M. S. & Mahmood, H. Short-Term Photovoltaic Power Forecasting Using an LSTM Neural Network and Synthetic Weather Forecast. *IEEE Access* **8**, 172524-172533 (2020). https://doi.org/10.1109/ACCESS.2020.3024901
38    Karevan, Z. & Suykens, J. A. K. Transductive LSTM for time-series prediction: An application to weather forecasting. *Neural Networks* **125**, 1-9 (2020). https://doi.org/https://doi.org/10.1016/j.neunet.2019.12.030

39  Abdalla, A. M., Ghaith, I. H. & Tamimi, A. A. Deep Learning Weather Forecasting Techniques: Literature Survey. in *2021 International Conference on Information Technology (ICIT).*  622-626. https://doi.org/10.1109/ICIT52682.2021.9491774
40  Kim, T.-Y. & Cho, S.-B. Predicting residential energy consumption using CNN-LSTM neural networks. *Energy* **182**, 72-81 (2019). https://doi.org/https://doi.org/10.1016/j.energy.2019.05.230
41  Le, T., Vo, M. T., Vo, B., Hwang, E., Rho, S. & Baik, S. W. Improving Electric Energy Consumption Prediction Using CNN and Bi-LSTM. *Applied Sciences* **9**, 4237 (2019).
42  Ullah, F. U. M., Ullah, A., Haq, I. U., Rho, S. & Baik, S. W. Short-Term Prediction of Residential Power Energy Consumption via CNN and Multi-Layer Bi-Directional LSTM Networks. *IEEE Access* **8**, 123369-123380 (2020). https://doi.org/10.1109/ACCESS.2019.2963045
43  Somu, N., Raman M R, G. & Ramamritham, K. A deep learning framework for building energy consumption forecast. *Renewable and Sustainable Energy Reviews* **137**, 110591 (2021). https://doi.org/https://doi.org/10.1016/j.rser.2020.110591
44  Khalil, M., McGough, A. S., Pourmirza, Z., Pazhoohesh, M. & Walker, S. Machine Learning, Deep Learning and Statistical Analysis for forecasting building energy consumption — A systematic review. *Engineering Applications of Artificial Intelligence* **115**, 105287 (2022). https://doi.org/https://doi.org/10.1016/j.engappai.2022.105287
45  Amalou, I., Mouhni, N. & Abdali, A. Multivariate time series prediction by RNN architectures for energy consumption forecasting. *Energy Reports* **8**, 1084-1091 (2022). https://doi.org/https://doi.org/10.1016/j.egyr.2022.07.139
46  Deng, C. & Liu, Y. A Deep Learning-Based Inventory Management and Demand Prediction Optimization Method for Anomaly Detection. *Wireless Communications and Mobile Computing* **2021**, 9969357 (2021). https://doi.org/https://doi.org/10.1155/2021/9969357
47  Seyedan, M., Mafakheri, F. & Wang, C. Order-up-to-level inventory optimization model using time-series demand forecasting with ensemble deep learning. *Supply Chain Analytics* **3**, 100024 (2023). https://doi.org/https://doi.org/10.1016/j.sca.2023.100024
48  Falatouri, T., Darbanian, F., Brandtner, P. & Udokwu, C. Predictive Analytics for Demand Forecasting – A Comparison of SARIMA and LSTM in Retail SCM. *Procedia Computer Science* **200**, 993-1003 (2022). https://doi.org/https://doi.org/10.1016/j.procs.2022.01.298
49  Joseph, R. V., Mohanty, A., Tyagi, S., Mishra, S., Satapathy, S. K. & Mohanty, S. N. A hybrid deep learning framework with CNN and Bi-directional LSTM for store item demand forecasting. *Computers and Electrical Engineering* **103**, 108358 (2022). https://doi.org/https://doi.org/10.1016/j.compeleceng.2022.108358
50  Abbasimehr, H., Shabani, M. & Yousefi, M. An optimized model using LSTM network for demand forecasting. *Computers & Industrial Engineering* **143**, 106435 (2020). https://doi.org/https://doi.org/10.1016/j.cie.2020.106435
51  Siami-Namini, S. & Namin, A. S. Forecasting economics and financial time series: ARIMA vs. LSTM. *arXiv preprint arXiv:1803.06386* (2018).
52  Cao, J., Li, Z. & Li, J. Financial time series forecasting model based on CEEMDAN and LSTM. *Physica A: Statistical Mechanics and its Applications* **519**, 127-139 (2019). https://doi.org/https://doi.org/10.1016/j.physa.2018.11.061
53  Tang, Y. *et al.* A survey on machine learning models for financial time series forecasting. *Neurocomputing* **512**, 363-380 (2022). https://doi.org/https://doi.org/10.1016/j.neucom.2022.09.003
54  Yamak, P. T., Yujian, L. & Gadosey, P. K. in *Proceedings of the 2019 2nd International Conference on Algorithms, Computing and Artificial Intelligence*   49–55 (Association for Computing Machinery, Sanya, China, 2020).
55  Pirani, M., Thakkar, P., Jivrani, P., Bohara, M. H. & Garg, D. A Comparative Analysis of ARIMA, GRU, LSTM and BiLSTM on Financial Time Series Forecasting. in *2022 IEEE International Conference on Distributed Computing and Electrical Circuits and Electronics (ICDCECE).*  1-6. https://doi.org/10.1109/ICDCECE53908.2022.9793213

56  Vaswani, A. *et al.* in *31st Conference on Neural Information Processing Systems* (Long Beach, CA, USA, 2017).
57  Wu, N., Green, B., Ben, X. & O'Banion, S. Deep transformer models for time series forecasting: The influenza prevalence case. *arXiv preprint arXiv:2001.08317* (2020).
58  Lara-Benítez, P., Gallego-Ledesma, L., Carranza-García, M. & Luna-Romera, J. M. Evaluation of the Transformer Architecture for Univariate Time Series Forecasting. in *Advances in Artificial Intelligence.* (eds Enrique Alba *et al.*) 106-115 (Springer International Publishing).
59  Sonata, I. & Heryadi, Y. Comparison of LSTM and Transformer for Time Series Data Forecasting. in *2024 7th International Conference on Informatics and Computational Sciences (ICICoS).* 491-495. https://doi.org/10.1109/ICICoS62600.2024.10636892
60  Nielsen, M. A. & Chuang, I. L. *Quantum Computation and Quantum Information: 10th Anniversary Edition*. (Cambridge University Press, 2010).
61  Preskill, J. Quantum Computing in the NISQ era and beyond. *Quantum* **2**, 79 (2018). https://doi.org/10.22331/q-2018-08-06-79
62  Arute, F. *et al.* Quantum supremacy using a programmable superconducting processor. *Nature* **574**, 505-510 (2019). https://doi.org/10.1038/s41586-019-1666-5
63  IBM. *IBM Quantum*, <https://quantum.ibm.com/> (
64  Computing, R. *Rigetti Quantum Cloud Services Documentation*, <https://docs.rigetti.com/qcs> (
65  IonQ. *IonQ Quantum Cloud*, <https://ionq.com/quantum-cloud> (
66  Xanadu. *Xanadu Quantum Technologies*, <https://www.xanadu.ai/> (
67  Honeywell. *Honeywell Quantum Solutions*, <https://www.honeywell.com/us/en/company/quantum> (
68  Fujitsu. *Fujitsu Quantum Computing Technologies*, <https://www.fujitsu.com/global/about/research/technology/quantum/index.html> (
69  Schuld, M. & Petruccione, F. *Supervised Learning with Quantum Computers*. (Springer International Publishing, 2018).
70  Biamonte, J., Wittek, P., Pancotti, N., Rebentrost, P., Wiebe, N. & Lloyd, S. Quantum machine learning. *Nature* **549**, 195-202 (2017). https://doi.org/10.1038/nature23474
71  Dunjko, V. & Briegel, H. Machine learning & artificial intelligence in the quantum domain: A review of recent progress. *Rep. Prog. Phys.* **81** (2018). https://doi.org/10.1088/1361-6633/aab406
72  Wang, Y. & Liu, J. A comprehensive review of quantum machine learning: from NISQ to fault tolerance. *Rep. Prog. Phys.* **87**, 116402 (2024). https://doi.org/10.1088/1361-6633/ad7f69
73  Peral-García, D., Cruz-Benito, J. & García-Peñalvo, F. J. Systematic literature review: Quantum machine learning and its applications. *Computer Science Review* **51**, 100619 (2024). https://doi.org/https://doi.org/10.1016/j.cosrev.2024.100619
74  Cerezo, M. *et al.* Variational quantum algorithms. *Nature Reviews Physics* **3**, 625-644 (2021). https://doi.org/10.1038/s42254-021-00348-9
75  Mitarai, K., Negoro, M., Kitagawa, M. & Fujii, K. Quantum circuit learning. *Phys. Rev. A* **98**, 032309 (2018). https://doi.org/10.1103/PhysRevA.98.032309
76  Chen, S. Y. C., Yoo, S. & Fang, Y. L. L. Quantum Long Short-Term Memory. in *ICASSP 2022 - 2022 IEEE International Conference on Acoustics, Speech and Signal Processing (ICASSP).* 8622-8626. https://doi.org/10.1109/ICASSP43922.2022.9747369
77  Lindsay, J. & Zand, R. A Novel Stochastic LSTM Model Inspired by Quantum Machine Learning. in *2023 24th International Symposium on Quality Electronic Design (ISQED).* 1-8. https://doi.org/10.1109/ISQED57927.2023.10129344
78  Cao, Y., Zhou, X., Fei, X., Zhao, H., Liu, W. & Zhao, J. Linear-layer-enhanced quantum long short-term memory for carbon price forecasting. *Quantum Machine Intelligence* **5**, 26 (2023). https://doi.org/10.1007/s42484-023-00115-2
79  Khan, S. Z., Muzammil, N., Ghafoor, S., Khan, H., Zaidi, S. M. H., Aljohani, A. J. & Aziz, I. Quantum long short-term memory (QLSTM) vs. classical LSTM in time series forecasting: a comparative study in solar power forecasting. *Frontiers in Physics* **12** (2024). https://doi.org/10.3389/fphy.2024.1439180

80. Gregory, J. W., Asai, K., Kameda, M., Liu, T. & Sullivan, J. P. A review of pressure-sensitive paint for high-speed and unsteady aerodynamics. *Proceedings of the Institution of Mechanical Engineers, Part G: Journal of Aerospace Engineering* **222**, 249-290 (2008). https://doi.org/10.1243/09544100jaero243
81. Peng, D. & Liu, Y. Fast pressure-sensitive paint for understanding complex flows: from regular to harsh environments. *Exp. Fluids* **61** (2020). https://doi.org/10.1007/s00348-019-2839-6
82. Westerweel, J., Elsinga, G. E. & Adrian, R. J. Particle Image Velocimetry for Complex and Turbulent Flows. *Annual Review of Fluid Mechanics* **45**, 409-436 (2013). https://doi.org/https://doi.org/10.1146/annurev-fluid-120710-101204
83. Huang, C.-Y., Matsuda, Y., Gregory, J. W., Nagai, H. & Asai, K. The applications of pressure-sensitive paint in microfluidic systems. *Microfluidics and Nanofluidics* **18**, 739-753 (2015). https://doi.org/10.1007/s10404-014-1510-z
84. Rost, S. & Thomas, C. ARRAY SEISMOLOGY: METHODS AND APPLICATIONS. *Rev. Geophys.* **40**, 2-1-2-27 (2002). https://doi.org/https://doi.org/10.1029/2000RG000100
85. Kiser, E. & Ishii, M. Back-Projection Imaging of Earthquakes. *Annual Review of Earth and Planetary Sciences* **45**, 271-299 (2017). https://doi.org/https://doi.org/10.1146/annurev-earth-063016-015801
86. Shiina, T., Maeda, T., Kano, M., Kato, A. & Hirata, N. An Optimum 2D Seismic‐Wavefield Reconstruction in Densely and Nonuniformly Distributed Stations: The Metropolitan Seismic Observation Network in Japan. *Seismological Research Letters* **92**, 2015-2027 (2021). https://doi.org/10.1785/0220200196
87. Xiao, X., Boles, S., Frolking, S., Li, C., Babu, J. Y., Salas, W. & Moore, B. Mapping paddy rice agriculture in South and Southeast Asia using multi-temporal MODIS images. *Remote Sensing of Environment* **100**, 95-113 (2006). https://doi.org/https://doi.org/10.1016/j.rse.2005.10.004
88. Gebbers, R. & Adamchuk, V. I. Precision Agriculture and Food Security. *Science* **327**, 828-831 (2010). https://doi.org/doi:10.1126/science.1183899
89. de Lima, I. P., Jorge, R. G. & de Lima, J. L. M. P. Remote Sensing Monitoring of Rice Fields: Towards Assessing Water Saving Irrigation Management Practices. *Frontiers in Remote Sensing* **2** (2021). https://doi.org/10.3389/frsen.2021.762093
90. Yang, M.-D., Tseng, H.-H., Hsu, Y.-C., Yang, C.-Y., Lai, M.-H. & Wu, D.-H. A UAV Open Dataset of Rice Paddies for Deep Learning Practice. *Remote Sensing* **13**, 1358 (2021).
91. Peitz, I. & van Leeuwen, R. Single-cell bacteria growth monitoring by automated DEP-facilitated image analysis. *Lab on a Chip* **10**, 2944-2951 (2010). https://doi.org/10.1039/C004691D
92. Kesavan, S. V. *et al.* High-throughput monitoring of major cell functions by means of lensfree video microscopy. *Scientific Reports* **4**, 5942 (2014). https://doi.org/10.1038/srep05942
93. Morishita, Y., Lee, S.-W., Suzuki, T., Yokoyama, H., Kamei, Y., Tamura, K. & Kawasumi-Kita, A. An archetype and scaling of developmental tissue dynamics across species. *Nature Communications* **14**, 8199 (2023). https://doi.org/10.1038/s41467-023-43902-y
94. Uchida, K. *et al.* Analysis of transonic buffet on ONERA-M4 model with unsteady pressure-sensitive paint. *Exp. Fluids* **62**, 134 (2021). https://doi.org/10.1007/s00348-021-03228-1
95. Muir, J. B. & Zhan, Z. Seismic wavefield reconstruction using a pre-conditioned wavelet–curvelet compressive sensing approach. *Geophysical Journal International* **227**, 303-315 (2021). https://doi.org/10.1093/gji/ggab222
96. Saito, Y., Nonomura, T., Yamada, K., Asai, K., Sasaki, Y. & Tsubakino, D. Determinant-based Fast Greedy Sensor Selection Algorithm. *IEEE Access* **9**, 68535-68551 (2021). https://doi.org/10.1109/ACCESS.2021.3076186
97. Nagata, T. *et al.* Seismic wavefield reconstruction based on compressed sensing using data-driven reduced-order model. *Geophysical Journal International* **233**, 33-50 (2022). https://doi.org/10.1093/gji/ggac443
98. Li, X., Hu, C., Liu, H., Shi, X. & Peng, J. Data-driven pressure estimation and optimal sensor selection for noisy turbine flow with blocked clustering strategy. *Phys. Fluids* **36** (2024). https://doi.org/10.1063/5.0239759

99  Jiao, L., Shi, Z., Wei, C., Ma, S., Wen, X., Liu, Y. & Peng, D. Fast pressure-sensitive paint measurements of dynamic stall on a pitching airfoil via intensity- and lifetime-based methods. *Journal of Visualization* **27**, 333-352 (2024). https://doi.org/10.1007/s12650-024-00973-3
100 Yokota, S. & Nonomura, T. Unsteady large-scale wake structure behind levitated free-stream-aligned circular cylinder. *J. Fluid Mech.* **982**, A10 (2024). https://doi.org/10.1017/jfm.2024.93
101 Wei, C., Zhang, H., Fan, H., Wang, P., Peng, D. & Liu, Y. Resolving high-frequency aeroacoustic noises of high-speed dual-impinging jets using fast pressure-sensitive paint. *Exp. Fluids* **65**, 131 (2024). https://doi.org/10.1007/s00348-024-03875-0
102 Sirovich, L. TURBULENCE AND THE DYNAMICS OF COHERENT STRUCTURES PART I: COHERENT STRUCTURES. *Quarterly of Applied Mathematics* **45**, 561-571 (1987).
103 Quarteroni, A., Manzoni, A. & Negri, F. *Reduced basis methods for partial differential equations: An introduction*. (2015).
104 Schmid, P. J. Dynamic mode decomposition of numerical and experimental data. *J. Fluid Mech.* **656**, 5-28 (2010). https://doi.org/10.1017/S0022112010001217
105 Kutz, J. N., Brunton, S. L., Brunton, B. W. & Proctor, J. L. *Dynamic Mode Decomposition: Data-Driven Modeling of Complex Systems*. (SIAM-Society for Industrial and Applied Mathematics, 2016).
106 Manohar, K., Brunton, B. W., Kutz, J. N. & Brunton, S. L. Data-Driven Sparse Sensor Placement for Reconstruction: Demonstrating the Benefits of Exploiting Known Patterns. *IEEE Control Systems Magazine* **38**, 63-86 (2018). https://doi.org/10.1109/MCS.2018.2810460
107 Inoue, T., Ikami, T., Egami, Y., Nagai, H., Naganuma, Y., Kimura, K. & Matsuda, Y. Data-driven optimal sensor placement for high-dimensional system using annealing machine. *Mechanical Systems and Signal Processing* **188** (2023). https://doi.org/10.1016/j.ymssp.2022.109957
108 Egami, Y., Hasegawa, A., Matsuda, Y., Ikami, T. & Nagai, H. Ruthenium-based fast-responding pressure-sensitive paint for measuring small pressure fluctuation in low-speed flow field. *Meas. Sci. Technol.* **32** (2021). https://doi.org/10.1088/1361-6501/abb916
109 Inoue, T., Matsuda, Y., Ikami, T., Nonomura, T., Egami, Y. & Nagai, H. Data-driven approach for noise reduction in pressure-sensitive paint data based on modal expansion and time-series data at optimally placed points. *Phys. Fluids* **33** (2021). https://doi.org/10.1063/5.0049071
110 National Renewable Energy, L. *Solar Power Data for Integration Studies*, <https://www.nrel.gov/grid/solar-power-data> (2006).
111 Caruana, R. Multitask Learning. *Machine Learning* **28**, 41-75 (1997). https://doi.org/10.1023/A:1007379606734
112 Crawshaw, M. Multi-task learning with deep neural networks: A survey. *arXiv preprint arXiv:2009.09796* (2020).
113 Liu, T., Sullivan, J. P., Asai, K., Klein, C. & Egami, Y. *Pressure and Temperature Sensitive Paints*. 2 edn, (Springer, Cham, 2021).
114 Bell, J. H., Schairer, E. T., Hand, L. A. & Mehta, R. D. Surface Pressure Measurements using Luminescent coatings. *Annual Review of Fluid Mechanics* **33**, 155-206 (2001). https://doi.org/doi:10.1146/annurev.fluid.33.1.155
115 Brunton, S. L. & Kutz, J. N. *Data-Driven Science and Engineering: Machine Learning, Dynamical Systems, and Control*. (Cambridge University Press, 2019).
116 Pastuhoff, M., Yorita, D., Asai, K. & Alfredsson, P. H. Enhancing the signal-to-noise ratio of pressure sensitive paint data by singular value decomposition. *Meas. Sci. Technol.* **24**, 075301 (2013). https://doi.org/10.1088/0957-0233/24/7/075301
117 Nonomura, T., Shibata, H. & Takaki, R. Extended-Kalman-filter-based dynamic mode decomposition for simultaneous system identification and denoising. *PLoS One* **14**, e0209836 (2019). https://doi.org/10.1371/journal.pone.0209836
118 Wen, X., Liu, Y., Li, Z., Chen, Y. & Peng, D. Data mining of a clean signal from highly noisy data based on compressed data fusion: A fast-responding pressure-sensitive paint application. *Phys. Fluids* **30** (2018). https://doi.org/10.1063/1.5046681

119  McKay, D. C. *et al.* Qiskit backend specifications for openqasm and openpulse experiments. *arXiv preprint arXiv:1809.03452* (2018).
120  Quantum, I. B. M. *Qiskit: An Open-source Framework for Quantum Computing*, <https://qiskit.org/> (2024).
121  Paszke, A. *et al.* Pytorch: An imperative style, high-performance deep learning library. *Advances in neural information processing systems* **32** (2019).
122  PyTorch, T. *PyTorch: An Open-source Machine Learning Framework*, <https://pytorch.org/> (2024).

# Supplementary Information

# Time-series forecasting for nonlinear high-dimensional system using hybrid method combining autoencoder and multi-parallelized quantum long short-term memory and gated recurrent unit

Makoto Takagi[1], Ryuji Kokubo[1], Misato Kurosawa[1], Tsubasa Ikami[2], Yasuhiro Egami[3], Hiroki Nagai[2], Takahiro Kashikawa[4], Koichi Kimura[4], Yutaka Takita[4], Yu Matsuda[1, *]

1. Department of Modern Mechanical Engineering, Waseda University, 3-4-1 Ookubo, Shinjuku-ku, Tokyo, 169-8555, Japan

2. Institute of Fluid Science, Tohoku University, 2-1-1 Katahira, Aoba-ku, Sendai, Miyagi-prefecture 980-8577, Japan

3. Department of Mechanical Engineering, Aichi Institute of Technology, 1247 Yachigusa, Yakusa-Cho, Toyota, Aichi-prefecture 470-0392, Japan

4. Quantum Application Core Project, Quantum Laboratory, Fujitsu Research, Fujistu Ltd, Kawasaki, Kanagawa 211-8588, Japan

* corresponding author: Yu Matsuda

Email: y.matsuda@waseda.jp

## Supplementary Note 1: Training and validation losses for pressure data obtained with GRU

The training and validation losses for pressure data using gated recurrent unit (GRU) are shown in Fig. S1. In the figure, the vertical axis represents the mean squared error (MSE). In this result, the training loss (MSE) approaches a steady value after 20 epochs, while the validation loss shows some fluctuations. The test loss (MSE) for GRU was $6.47 \times 10^{-2}$, which was higher than that of $6.39 \times 10^{-2}$ obtained with the multi-parallelized quantum GRU (MP-QGRU).

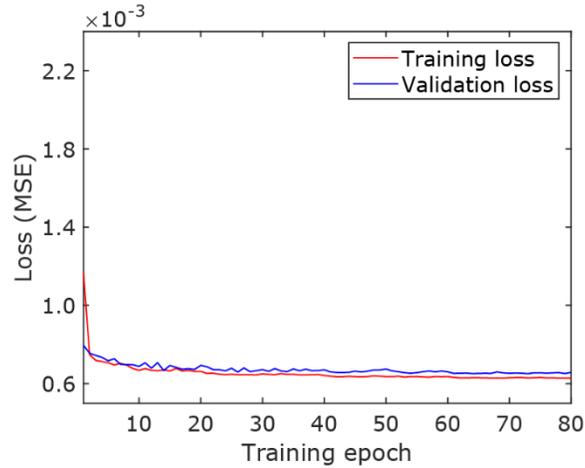

**Figure S1** Training and validation loss for pressure data using GRU

## Supplementary Note 2: Time-series forecasting results of GRU and L-QLSTM

The time-series forecasting results of GRU and linear-layer-enhanced quantum long-short term memory (L-QLSTM) are shown in Fig. S2. Both results show good agreement with each other. Also in comparison with Fig. 4 in the main text, the results also align well with that obtained using LSTM. Similar to LSTM, both L-QLSTM and GRU exhibited a tendency to underestimate local minima, especially around time step 60.

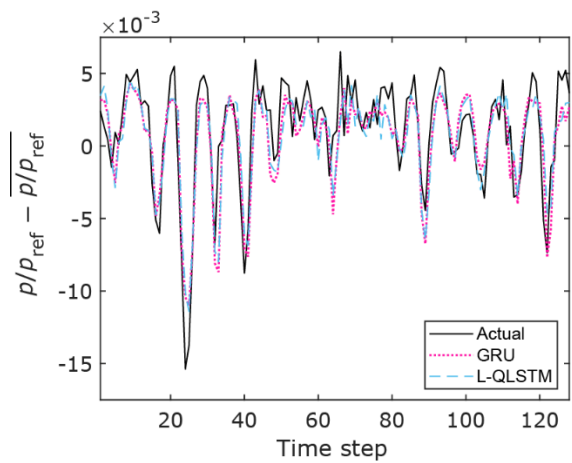

**Figure S2** Time-series forecasting results for pressure data. Forecasting results of GRU and L-QGRU are shown.

**Supplementary Note 3: Time-series forecasting for solar power data**

The performance of MP-QLSTM and MP-QGRU was compared with those of LSTM, GRU, and L-QLSTM using solar power data provided by the National Renewable Energy Laboratory (NREL), [1] which were used to evaluate the performance of QLSTM in literature. [2] As the performance of the original QLSTM has been reported to be worse than that of LSTM and L-QLSTM, [3] this study focused on comparisons between LSTM, L-QLSTM, and our proposed MP-QLSTM and MP-QGRU. In this study, we implemented the calculations through a script developed using Qiskit [4,5] and PyTorch. [6,7] We used the preprocessed data provided by the authors of Reference, [2] prepared using the same method as described in their study. The dataset was constructed by integrating weather data (such as temperature, humidity, and wind speed) recorded at 30-minute intervals and solar power generation data recorded every 5 minutes, aligning them to create 10-minute interval data, and normalized for training. The weather data were linearly interpolated to match the 10-minute resolution. This dataset contains 52,558 data points recorded every 10 minutes from January 1, 2006, 00:00 to December 31, 2006, 23:30. Of these, the first 70% were used as training data, the following 20% as validation data, and the remaining 10% as test data. In this study, the dimensions of both the input data and hidden layer were set to 9, and the window length was fixed to 8 for each method. In the calculation by MP-QLSTM and MP-QGRU, both the number of linear layers for VQC, $M$, and the number of qubits for each VQC were set to 3. The training was performed using the Adaptive moment estimation (Adam) optimizer of "torch.optim" with a learning rate of 0.001, and the batch size was set to 32. For the L-QLSTM calculation, the number of qubits in the VQCs was set to 4.

The test losses, evaluated using the mean absolute error (MAE), were $1.38 \times 10^{-2}$ MW for MP-QLSTM, $1.46 \times 10^{-2}$ MW for MP-QGRU, $1.47 \times 10^{-2}$ MW for LSTM, $1.41 \times 10^{-2}$ MW for GRU, and $1.72 \times 10^{-2}$ MW for L-QLSTM. The test losses, evaluated using the mean squared error (MSE), were $1.29 \times 10^{-3}$ (MW)$^2$ for MP-QLSTM, $1.33 \times 10^{-3}$ (MW)$^2$ for MP-QGRU, $1.34 \times 10^{-3}$ (MW)$^2$ for LSTM, $1.33 \times 10^{-3}$ (MW)$^2$ for GRU, and $1.43 \times 10^{-3}$ (MW)$^2$ for L-QLSTM. The test loss of MP-QLSTM was approximately 6%(MAE) and 3%(MSE) lower than those of LSTM and GRU and approximately 24%(MAE) and 10%(MSE) lower than that of L-QLSTM. The time-series forecasting results of LSTM, MP-QLSTM, and MP-QGRU are shown in Fig. S3. As a typical example, the figure shows the solar power generation from 4:00 (before sunrise) to 20:00 (after sunset) on selected days. As shown in Fig. S3(a), all methods show very good agreement with the actual data for cases where the solar power generation changes smoothly. Notably, MP-QLSTM provides the best performance in this case. In contrast, as can be seen around 12:30 in Fig. S3(b), all methods show a certain delay to a sudden drop in power generation. These indicate that the predictions are generally accurate except in cases of abrupt changes.

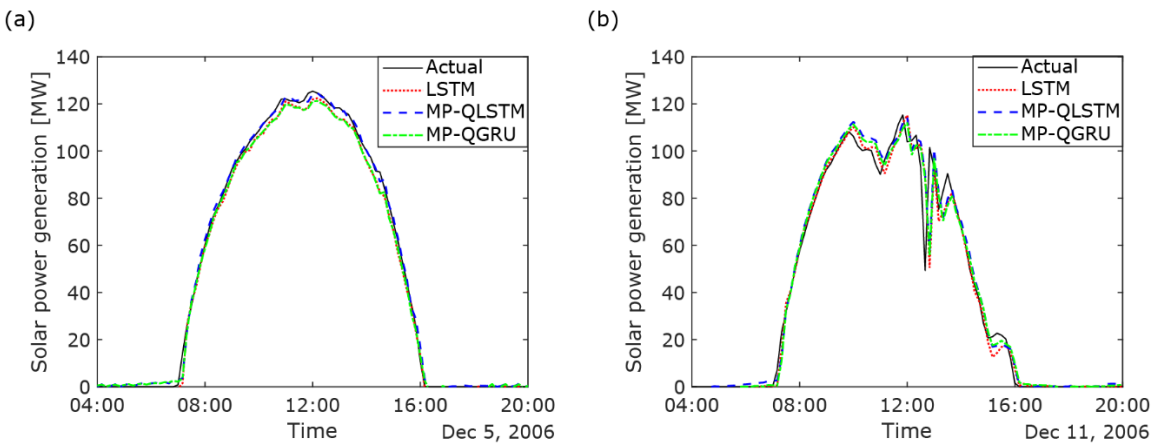

**Figure S3** Time-series forecasting results for solar power data provided by NREL. Forecasting results of LSTM, MP-QLSTM, and MP-QGRU for (a) data from December 5, 2006, and (b) data from December 11, 2006.

**Supplementary Note 4: Time-series forecasting for Lorenz equations**

Based on the above discussion and the results presented in the main text, it has been shown that MP-QLSTM and LSTM provide better performance compared to the other methods. Therefore, a more detailed comparison between MP-QLSTM and LSTM was conducted using data generated by the Lorenz equations with added Gaussian noise. The Lorenz equations are given as follows:

$$\frac{dx}{dt} = \sigma(y - x), \tag{1}$$

$$\frac{dy}{dt} = rx - y - xz, \tag{2}$$

$$\frac{dz}{dt} = xy - bz, \tag{3}$$

where $x$, $y$, and $z$ are time dependent variables, and $\sigma, r$, and $b$ are parameters. In this study, we set the parameters as $\sigma = 10$, $r = 28$, and $b = 8/3$. Following the literature, [8] the numerical solutions of the equations were obtained under the initial conditions of $x(0) = 0$, $y(0) = -0.01$, and $z = 9$, with a time step of $\Delta t = 0.01$. We generated the 5000 time step data, and used the first 70% data as training data, the following 20% as validation data, and the remaining 10% as test data. The noisy data was generated by adding Gaussian noise with a mean of 0 and standard deviations (std) of 1, 3, and 5 to the original data. Then, the data was normalized for training. The dimensions of both the input data and hidden layer were set to 3, and the window length was fixed to 20 for both methods. In the calculation by MP-QLSTM, both the number of linear layers for VQC, $M$, and the number of qubits for each VQC were set to 3. The training was performed using the Adam optimizer with a learning rate of 0.001, and the batch size was set to 64. Each method was performed five times, and the mean and standard deviation of MAEs and MSEs for test losses were also calculated to evaluate variability. The results are presented in Table S1. As a typical results, the forecasting results for $x$, $y$, and $z$ for the noisy data of std = 5 are shown in Fig. S4. As shown in this Table, while the test losses of MP-QLSTM are larger than those of LSTM for the noisy data of std = 1 and std 3, test losses of MP-QLSTM are smaller than those of LSTM for the noisy data of std = 5. The forecasting results of MP-QLSTM is smoother than those of LSTM, as shown in Fig. S4. For example, in Fig. S4 (c), LSTM appears to overreact to noise, as seen in its response to the spike-like fluctuation around Time = 47.5. From this perspective, MP-QLSTM is more suitable for forecasting data with substantial noise. On the other hand, when the noise level is low, LSTM performs better, suggesting that it is important to choose between the two methods depending on the characteristics of the data.

Table S1  MAEs and MSEs for test losses of MP-QLSTM and LSTM. The mean values and standard deviations are shown.

|  | Gaussian (std = 1) | | Gaussian (std = 3) | | Gaussian (std = 5) | |
| --- | --- | --- | --- | --- | --- | --- |
|  | MAE | MSE | MAE | MSE | MAE | MSE |
| MP-QLSTM | $1.20 \pm 0.20$ | $2.52 \pm 0.80$ | $2.87 \pm 0.20$ | $13.1 \pm 1.9$ | $4.34 \pm 0.04$ | $29.2 \pm 1.0$ |
| LSTM | $0.99 \pm 0.12$ | $1.65 \pm 0.39$ | $2.72 \pm 0.12$ | $11.5 \pm 1.0$ | $4.48 \pm 0.15$ | $30.7 \pm 1.5$ |

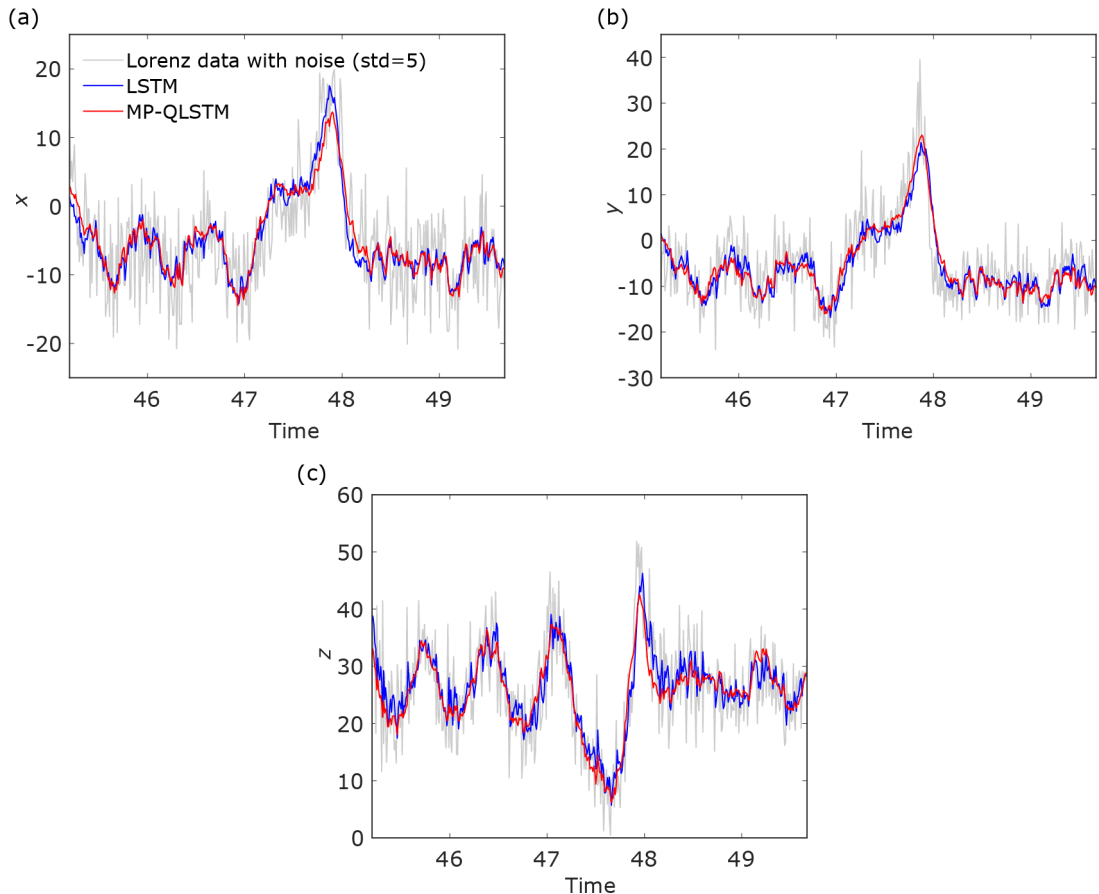

**Figure S4** Time-series forecasting results for the Lorenz equations with noise. Forecasting results of (a) $x$, (b) $y$, and (c) $z$.

**References**

1   National Renewable Energy, L. *Solar Power Data for Integration Studies*, <https://www.nrel.gov/grid/solar-power-data> (2006).
2   Khan, S. Z., Muzammil, N., Ghafoor, S., Khan, H., Zaidi, S. M. H., Aljohani, A. J. & Aziz, I. Quantum long short-term memory (QLSTM) vs. classical LSTM in time series forecasting: a comparative study in solar power forecasting. *Frontiers in Physics* **12** (2024). https://doi.org/10.3389/fphy.2024.1439180
3   Cao, Y., Zhou, X., Fei, X., Zhao, H., Liu, W. & Zhao, J. Linear-layer-enhanced quantum long short-term memory for carbon price forecasting. *Quantum Machine Intelligence* **5**, 26 (2023). https://doi.org/10.1007/s42484-023-00115-2
4   McKay, D. C. *et al.* Qiskit backend specifications for openqasm and openpulse experiments. *arXiv preprint arXiv:1809.03452* (2018).
5   Quantum, I. B. M. *Qiskit: An Open-source Framework for Quantum Computing*, <https://qiskit.org/> (2024).
6   Paszke, A. *et al.* Pytorch: An imperative style, high-performance deep learning library. *Advances in neural information processing systems* **32** (2019).
7   PyTorch, T. *PyTorch: An Open-source Machine Learning Framework*, <https://pytorch.org/> (2024).
8   Rivera-Ruiz, M. A., Mendez-Vazquez, A. & López-Romero, J. M. Time Series Forecasting with Quantum Machine Learning Architectures.   66-82 (Springer Nature Switzerland).